\documentclass[%
aip,
amsmath,amssymb,
 reprint,%
]{revtex4-1}

\usepackage{graphicx}
\usepackage{dcolumn}
\usepackage{bm}
\usepackage[dvipsnames]{xcolor}
\usepackage[utf8]{inputenc}
\usepackage[T1]{fontenc}
\usepackage{mathptmx}
\usepackage{etoolbox}
\usepackage{xfrac}

\usepackage{soul}
\usepackage[none]{hyphenat}

\makeatletter
\def\@email#1#2{%
 \endgroup
 \patchcmd{\titleblock@produce}
  {\frontmatter@RRAPformat}
  {\frontmatter@RRAPformat{\produce@RRAP{*#1\href{mailto:#2}{#2}}}\frontmatter@RRAPformat}
  {}{}
}%
\makeatother

\begin{document}

\preprint{AIP/123-QED}

\title[]{Machine learning approach to detect dynamical states from recurrence measures\\}

\author{Dheeraja Thakur}
\affiliation{School of Physics, Indian Institute of Science Education and Research, Thiruvananthapuram, 695551, Kerala, India.}
\author{Athul Mohan}
 
\affiliation{ Mahatma Gandhi University, Kottayam. 686560, Kerala, India.}

\author{G Ambika}
 
\affiliation{School of Physics, Indian Institute of Science Education and Research, Thiruvananthapuram, 695551, Kerala, India.}
\author{Chandrakala Meena}
 \email{c.meena@iisertvm.ac.in}
\affiliation{School of Physics, Indian Institute of Science Education and Research, Thiruvananthapuram, 695551, Kerala, India.}

\date{\today}

\begin{abstract}
We integrate machine learning approaches with nonlinear time series analysis, specifically utilizing recurrence measures to classify various dynamical states emerging from time series. We implement three machine learning algorithms: Logistic Regression, Random Forest, and Support Vector Machine for this study. The input features are derived from the recurrence quantification of nonlinear time series and characteristic measures of the corresponding recurrence networks. For training and testing we generate synthetic data from standard nonlinear dynamical systems and evaluate the efficiency and performance of the machine learning algorithms in classifying time series into periodic, chaotic, hyperchaotic, or noisy categories. Additionally, we explore the significance of input features in the classification scheme and find that the features quantifying the density of recurrence points are the most relevant. Furthermore, we illustrate how the trained algorithms can successfully predict the dynamical states of two variable  stars, SX Her and AC Her from the data of their light curves. We also indicate how the algorithms can be trained to classify data from discrete systems.
\end{abstract}

\maketitle

\begin{quotation}
The application of techniques based on the recurrences in dynamical systems using recurrence plot based  methods, such as recurrence quantification analysis and recurrence networks, is of growing interest in many different scientific disciplines. One specific advantage of this approach compared to purely statistical approaches is that these techniques bring out underlying dynamical details and characteristics that may not be apparent otherwise. The suggested methods in these cases can be easily tailored to fit any relevant requirements like diagnosis, classification, control and compression of data and signals. The power of such measures in identifying different dynamical states like periodic, chaotic, hyperchaotic or random variations based on available data from complex systems is being tested in specific case studies.

We present an exhaustive and exploratory study on the use of various recurrence measures as input features in predicting different dynamical states like periodic, chaotic and hyperchaotic that emerge from a given time series and differentiating them from noisy or stochastic data. We investigate the role of the recurrence threshold by considering two choices, with a fixed recurrence rate of $0.1$ and $0.5$ in standard deviation units.  We generate time series data from standard dynamical continuous systems in different dynamical states and apply three supervised machine learning algorithms, namely, Logistic Regression, Random Forest, and Support Vector Machine. We estimate their efficiency to classify multiple dynamical states associated with the nonlinear time series, in terms of average accuracy as well as class-specific accuracy, precision, sensitivity, specificity, and F1 score. 

Our analysis reveals that, on an average, the three classifiers perform well. However, Random Forest in particular, exhibits higher accuracy than Support Vector Machine and Logistic Regression in classifying the multiple dynamical states underlying the nonlinear time series. We also find that random forest and support vector machine classifiers can tolerate noise contamination in data up to \textbf{5\%}. Moreover, we observe the relative importance of various input features in determining the dynamical states accurately. We demonstrate how the algorithms, trained on nonlinear time series generated from synthetic models, correctly predict the dynamical states of two variable stars from their time series data.Further we extend our analysis for the data generated from discrete systems.Thus, the algorithms, developed with proper training using a large amount of data and incorporating built-in codes for extracting relevant features from recurrence measures, provide a unique and efficient tool for determining dynamical states from time series data.

\end{quotation}

\section{\label{sec:level1}Introduction}
Time series analysis is an indispensable field of recent research involving large and multiple data sets with far-reaching implications across various disciplines, including finance \cite{Finance}, economics \cite{Finu}, climate science \cite{ClimateScience}, healthcare \cite{Healthcare} and beyond. It encompasses the study of features collected sequentially over time, offering a dynamic perspective on trends, patterns, and dependencies within the data.  In this context, nonlinear time series analysis represents an essential facet of analysis, offering profound insights into complex systems across diverse domains \cite{astronomy, Biology, Neuroscience}.  Unlike traditional linear models, which assume simple relationships, nonlinear time series analysis delves into the intricate, dynamic interactions within sequential data, shedding light on hidden patterns and emergent phenomena. These techniques are driven by the recognition that many real-world processes exhibit nonlinear behavior, rendering linear models inadequate for capturing the complexities inherent in the data. The Lyapunov exponent serves as a potent tool for assessing the dynamical state of a system by analyzing its dynamical equations. However, its utility is not universal, as it may not always be applicable due to either the absence of accurate models \cite{2DHLM} or the inadequacy of models in capturing the correct dynamics of real systems. Numerous studies are underway to develop more broadly applicable tools for detecting dynamical states. For example, embedding techniques have emerged, which refer to the process of transforming a one-dimensional time series into a higher-dimensional phase space representation. This transformation enables the exploration of hidden dynamics and intricate dependencies that may not be discernible in the original data.  \\
\indent Recurrence is a natural characteristic of bounded dynamical systems and found to occur in numerous real world systems in atmospheric science \cite{ClimateChange}, astrophysics \cite{Binary}, and physiological systems \cite{Medicalu}. We note that various dynamical behaviours like periodic, chaotic and hyperchaotic as well as noisy patterns are observed in many systems from climate  \cite{meena2017effect, Meena2017CoexistenceOA} to biology \cite{baleanu2021hyperchaotic,meena2017threshold}. The extent to which a system repeats its patterns speaks a great deal about its underlying dynamics. The techniques in recurrence analysis explore the temporal dependencies and patterns within a time series by identifying and quantifying instances when the system revisits or recurs near to a previously encountered state in its phase space. This is achieved by constructing the recurrence plot, which is a pictorial representation of the recurrences in the trajectory reconstructed from the data  \cite{interpretation}. \\
\indent Machine learning algorithms play a highly relevant role in modern technology and address automation problems in many fields as these techniques can be used to identify features with high sensitivity. Over the past decade, many studies are reported that implement machine learning models using existing data repositories and investigate the idea and feasibility of introducing a generalized classifier.  The development of such machine learning(ML) techniques helps to automate complex manual programs. In addition, the availability, scale, and complexity of datasets demands the need for faster, more accurate, and more reliable automation methods for extracting information, reforming, pre-processing, and analyzing them in the most effective ways. With the advent of machine learning, many attempts are being made to extract features from the time series that can be used for the binary class classification \cite{EEG,ML} and multi-class classification \cite{gawade2022artificial} methods. Recently, numerous studies have emerged combining machine learning approaches with network analysis to predict diverse patterns within complex systems\cite{lopes2022machine, pessa2022determining, deepcriminal}. It is reported that the channel-frequency convolutional neural network (CFCNN), combined with recurrence quantification analysis (RQA), helps in the robust recognition of electroencephalogram (EEG) signals collected from different emotion states \cite{25}. Also, RQA, with the well-known machine learning algorithm like support vector machine, is useful for the binary classification of protein sequences\cite{26}. An effective paroxysmal atrial fibrillation predictor, which is based on the analysis of the RR-interval signal using recurrence plot-based features, is also reported \cite{27}.While these approaches establish the efficiency of RQA measures, each is developed for specific case studies employing one or two ML techniques. \\
\indent The methods in data analysis employ machine learning algorithms, mostly using statistical measures for detection, control and predictions in studies related to climate, astrophysics etc. \cite{Mapping,19,20,21,22,23,24}. However, recent trends indicate that a more effective classification is possible by considering the dynamical features of data.  Among them, the most powerful are the ones that combine recurrence-based measures with machine learning \cite{MLAndTimeseries}.
In the present work,  we report a detailed and exhaustive study on the classification of dynamical states using three different ML techniques.\\
\indent We start with the synthetic time series data generated to encompass periodic, chaotic, hyperchaotic, and white noise patterns. We generate synthetic data from the standard nonlinear continuous dynamical systems. The input features for the machine learning algorithms are derived from the recurrence quantification of nonlinear time series and characteristic measures of the corresponding recurrence networks.   We implement three machine learning algorithms:
Logistic Regression, Random Forest, and Support Vector Machine for this study and evaluate their
efficiency and performance in classifying time series into periodic, chaotic, hyperchaotic, or noisy categories. We choose the recurrence threshold in two different ways and check their efficiency.   We also study the relative importance of all the features in identifying and predicting the dynamic state from the given data and deduce that the features quantifying recurrence points are the most important ones in classification.  Furthermore, we illustrate how
the trained algorithms can successfully predict the dynamical states of two variable stars, SX Her and AC Her, from the
data of their light curves relying on RQA measures and Recurrence Network (RN) characteristics as features for classification.We also indicate how the same machine learning algorithms can be trained to classify data from discrete dynamical systems.

In the following sections, we present the methodology, classification algorithms and the main results of this study. Our conclusions are summarized in the last section.

\begin{table*}
\caption{\label{tab:table1_models} Details of continuous dynamical systems and their parameter values used to generate synthetic data with dynamical states of periodic, chaotic hyperchaotic and noisy behaviours.}
\begin{ruledtabular}
\begin{tabular}{p{1.5cm}p{3.5cm}p{3.5cm}p{0.5cm}p{2.5cm}}
 Time Series Generator & Dynamical Equations & Parameter Values & & Dynamical Behavior\\\hline
Lorenz System \cite{article} & 
    $\begin{aligned}
        \dot{x} &= \sigma(y - x) \\
        \dot{y} &= x(\rho - z) - y \\
        \dot{z} &= xy - \beta z
    \end{aligned}$ &
    $\begin{aligned} 
        \sigma &= 10, \quad
        \beta = 8/3, \quad
        \rho \in (28, 85) \\
        \sigma &= 10, \quad
        \beta = 8/3, \quad
        \rho \in (99.6, 99.8) 
    \end{aligned}$ & &
    $\begin{aligned} 
        \text{Chaotic} \\
        \text{Periodic}
    \end{aligned}$ \\\\ \hline
R{\"o}ssler System \cite{ibrahim2018chaotic} & 
    $\begin{aligned}
        \dot{x} &= -y - z \\
        \dot{y} &= x + a y \\
        \dot{z} &= b + z(x - c)
    \end{aligned}$ &
    $\begin{aligned} 
        a &= 0.2, \quad
        b = 0.2, \quad
        c \in (5, 7.5) \\
        a &= 0.2, \quad
        b = 0.2, \quad
        c \in (0.8, 4.1) 
    \end{aligned}$ & &
    $\begin{aligned} 
        \text{Chaotic} \\
        \text{Periodic}
    \end{aligned}$ \\\\ \hline 
Duffing System \cite{duffingnew} & 
    $\begin{aligned}
        \dot{x} &= y \\
        \dot{y} &= x - x^3 - \delta y + a \sin(\omega t)
    \end{aligned}$ &
    $\begin{aligned} 
        \delta &= 0.5, \quad
        \omega = 1, \quad
        a \in (0.6, 0.8) \\
        \delta &= 0.5, \quad
        \omega = 1, \quad
        a \in (0, 0.35) 
    \end{aligned}$ & &
    $\begin{aligned} 
        \text{Chaotic} \\
        \text{Periodic}
    \end{aligned}$ \\\\ \hline
Chen System \cite{CHEN2007696} & 
    $\begin{aligned}
        \dot{x} &= a(y - x) + eyz \\
        \dot{y} &= cx - dxz + y + u \\
        \dot{z} &= xy - bz \\
        \dot{w} &= -ky
    \end{aligned}$ &
    $\begin{aligned} 
        &a= 35, \quad b= 4.9, \quad c = 25, \quad d = 5, \quad e = 35 \\  
        &k \in (0, 171), (241,297)   
    \end{aligned}$ & &
    $\begin{aligned} 
        \text{Hyperchaotic} 
    \end{aligned}$ \\\\ \hline
4D Lorenz System \cite{LorenzHyper} & 
    $\begin{aligned}
        \dot{x} &= a(y - x)  \\
        \dot{y} &= bx - xz - cy + w \\
        \dot{z} &= xy - dz \\
        \dot{w} &= -ky-rw
    \end{aligned}$ &
    $\begin{aligned} 
        &a= 12, \quad b= 23, \quad c = 1, \quad d = 2.1, \quad r = 0.2 \\  
        &k \in (2.35,8.03), (12.80,15.69)   
    \end{aligned}$ & &
    $\begin{aligned} 
        \text{Hyperchaotic} 
    \end{aligned}$ \\\\ \hline
4D R{\"o}ssler System \cite{RosslerHyper} & 
    $\begin{aligned}
        \dot{x} &= -y-z  \\
        \dot{y} &= x+ay+w \\
        \dot{z} &= b+xz \\
        \dot{w} &= -cz+dw
    \end{aligned}$ &
    $\begin{aligned} 
        &a= 0.25, \quad c = 0.5, \quad d = 0.05 \\  
        &b \in (3,5)   
    \end{aligned}$ & &
    $\begin{aligned} 
        \text{Hyperchaotic} 
    \end{aligned}$ \\\\ \hline
Gaussian distribution & 
    $\begin{aligned} 
    X \sim N(\mu, \sigma^2) 
    \end{aligned}$ & $
    \begin{tabular}[t]{@{}l@{}}$\mu$ =0\\ $\sigma$=(0,100) \end{tabular}  $ & &
    $\begin{aligned} 
        \text{White noise} 
    \end{aligned}$ \\
        \hline

\end{tabular} 
\end{ruledtabular}
\end{table*}

\section{\label{sec:level2}Methodology}
We focus on the multi-class classification of nonlinear time series into four distinct classes: periodic, chaotic, hyperchaotic and noise.

 We first generate a set of time series data (FIG.\ref{Illustrative}(a)) from synthetic models, including standard nonlinear systems like Lorenz\cite{article}, R{\"o}ssler\cite{ibrahim2018chaotic}, Duffing\cite{meena2020resilience}, and Chen\cite{CHEN2007696}, 4D R{\"o}ssler \cite{RosslerHyper} and 4D Lorenz \cite{LorenzHyper} systems. We then use the x-variable time series for the reconstruction of their trajectory or attractor (FIG.\ref{Illustrative}(b)) in higher dimensions employing Taken's embedding. Then, we construct the corresponding recurrence plots ((FIG.\ref{Illustrative}(c)) and recurrence networks ((FIG.\ref{Illustrative}(d)). The essential features for machine learning algorithms are extracted through recurrence quantification analysis of the recurrence plots and average network characteristics derived from the recurrence networks (FIG.\ref{Illustrative}(e)).

Next, we gather labeled datasets for machine learning algorithms, setting aside $80\%$ of the data for training and allocating $20\%$ for prediction. We employ three machine learning algorithms as classifiers (FIG.\ref{Illustrative}(f)) and assess their performance.
 In the following subsections, we present a brief overview of the methods for reconstructing attractors from a single nonlinear time series, creating recurrence plots and networks, and extracting features. We also provide the methodology of the three machine learning algorithms for the multi-class classification of the dynamical states, namely, periodic, chaotic, hyperchaotic and noise (Fig.\ref{Illustrative}(g)).
\subsection{Phase space reconstruction from time series}
We reconstruct the higher dimensional phase space trajectory of each time series, applying Taken's embedding technique\textcolor{red}{\cite{Taken}}. The reconstruction process involves the creation of $m$ dimensional vectors from a single time series $u(t_i)$. These points separated by delay time '$\tau$' \cite{Ambika2019MethodsON} 
form the vectors representing the point at a given time in the reconstructed $m$ dimensional phase space. 
\begin{equation}
x_i = [u(t_i), u(t_i + \tau), \ldots, u(t_i + (m-1)\tau)]
\end{equation}
 Following the standard procedure, the delay parameter '$\tau$' is chosen as the time when the autocorrelation function first decreases to  $1/e$ of its value \cite{autocorrelation}. To find the embedding dimension, we consider the false nearest neighbour (FNN) approach \cite{FNN}. This method iteratively examines the fraction of FNNs for increasing embedding dimension and finds the minimum embedding dimension $m$, where this fraction goes to zero. 
\subsection{Recurrence Plots and Recurrence Networks}
The construction of a recurrence plot looks for pairwise closeness of all possible pairs of points on the reconstructed attractor. 
Basically, the state of the system is compared using this tool at two different times, $i$ and $j$, to construct a binary and symmetric matrix $R$ \cite{eckmann1995recurrence}, where $R_{ij}$ = 1 if the state $x_j$ is a neighbour of $x_i$ within the chosen threshold $\epsilon$ in phase space, and $R_{ij}$ = 0 otherwise \cite{RecurrenceMatrix}.  Different methods are available for fixing the value of the threshold $\epsilon$ \cite{interpretation, pitfalls}. We follow two methods in our analysis to choose its value; the value such that the recurrence rate is  0.1 \cite{interpretation}(henceforth $\epsilon_1$), and the value of 0.5 in standard deviation units  \cite{stdev} (henceforth $\epsilon_2$).
\noindent Recurrence network is defined by the adjacency matrix derived from the recurrence matrix using 
$A_{ij}(\epsilon) = R_{ij}(\epsilon) - \delta_{ij}$ where $\delta_{ij}$ is the identity matrix of the same dimensions that helps to remove the self-loop within the network\cite{yunicorn}.

\subsection{Features Extraction}
From the recurrence plots constructed as described in the above section, we compute measures relevant for feature identification. RQA is used for quantifying different structures such as dots, diagonal and vertical lines, in a recurrence plot. Here, we consider six statistically significant recurrence measures \cite{best} for the quantification of recurrence plots. These features are extracted using PyUnicorn package\cite{yunicorn} in python. 
\\

\noindent Recurrence Rate(RR) is the measure of the density of recurrence points. It excludes the line of identity from the calculation \cite{Extra1}. 
\begin{equation}
RR = \frac{1}{N^2} \sum_{i,j=1}^{N} R_{i,j}
\end{equation}
Determinism(DET) gives the density of recurrence points forming diagonal lines\cite{pitfalls} 
\begin{equation}
DET = \frac{\sum_{l=l_{\text{min}}}^{N} l  P(l)}{\sum_{l=1}^{N} l  P(l)}
\end{equation}
where 
$P(l)$ is the frequency distribution of the lengths $l$ of the diagonal lines and $l_{\text{min}}$ is the least length considered. 
Similarly, the density of recurrence points forming vertical lines is given by Laminarity(LAM) 
\begin{equation}
\text{LAM} = \frac{\sum_{v=v_{\min}}^{N} vP(v)}{\sum_{v=1}^{N} vP(v)}
\end{equation}
where $P(v)$ is the frequency distribution of the lengths $v$ of the vertical lines and$v_{\text{min}}$ is the minimum length considered.
The length of the longest diagonal line is given by $L_{\text{max}}$
\begin{equation}
L_{\text{max}}
 = \max\left\{ l_i \right\}_{i=1}^{N_l}
\end{equation}
where $N_l = \sum_{l \geq l_{\text{min}}} P(l)$ is the total number of diagonal lines.
Shannon entropy (ENTR) gives the entropy of the diagonal lines in the recurrence plot as
\begin{equation}
ENTR = - \sum_{l=l_{\text{min}}}^{N} P(l) \ln P(l)
\end{equation}
 Another important measure which calculates the average time the system remains in the given state is Trapping Time(TT). 
\begin{equation}
TT = \frac{\sum_{v=v_{\text{min}}}^{N} vP(v)}{\sum_{v=v_{\text{min}}}^{N} P(v)}
\end{equation}
We also consider two measures from the recurrence networks which are computed as follows:\\
\noindent Average path length (L) gives the average of the shortest path between all possible pairs of the nodes of a network\cite{avgPathLength}.
\begin{equation}
CPL =  \frac{1}{N(N-1)} \sum_{ij=1, i \neq j}^{N} d_{ij}
\end{equation}
where N is the number of nodes and $d_{ij}$ is the shortest path between nodes $i$ and $j$.\\
Global clustering coefficient (GCC) is the average of local clustering coeffients \(C_i\)\cite{GCC}
\begin{equation}
GCC =\frac {\sum_{i=1}^{N}{C_i}}{N}
\end{equation}
where $C_i$ is the local clustering coefficient of the i$^{th}$ node of degree $k_i$.
\begin{equation}
C_i = \frac{\sum_{jk}A_{ij}A_{jk}A_{ki}}{k_i (k_{i}-1)
}
\end{equation}

\begin{figure*}
  \centering
  \includegraphics[width=\textwidth]{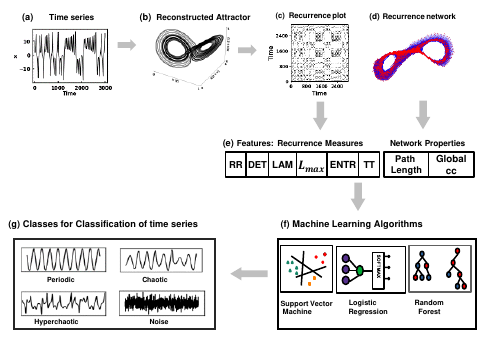}
  \caption{\label{Illustrative}Schematic of the methodology developed for the classification of the time series data using recurrence quantification and machine learning algorithms. From the given (a) nonlinear time series, we reconstruct the attractor (b) using time delay embedding. Following this, we generate (c) recurrence plots and (d) recurrence networks (nodes are shown in red color and edges are shown in blue color). Then, we extract \textcolor{red}{(e)} six recurrence measures from the recurrence plots and two measures from the recurrence networks. These measures serve as features for the machine learning algorithms (f) Logistic Regression, Random Forest and Support Vector Machine. The final step(g) classifies the time series into four classes: periodic, chaotic, hyperchaotic, or noise.}
\end{figure*}
\begin{figure}
\includegraphics[width=0.5\textwidth]
{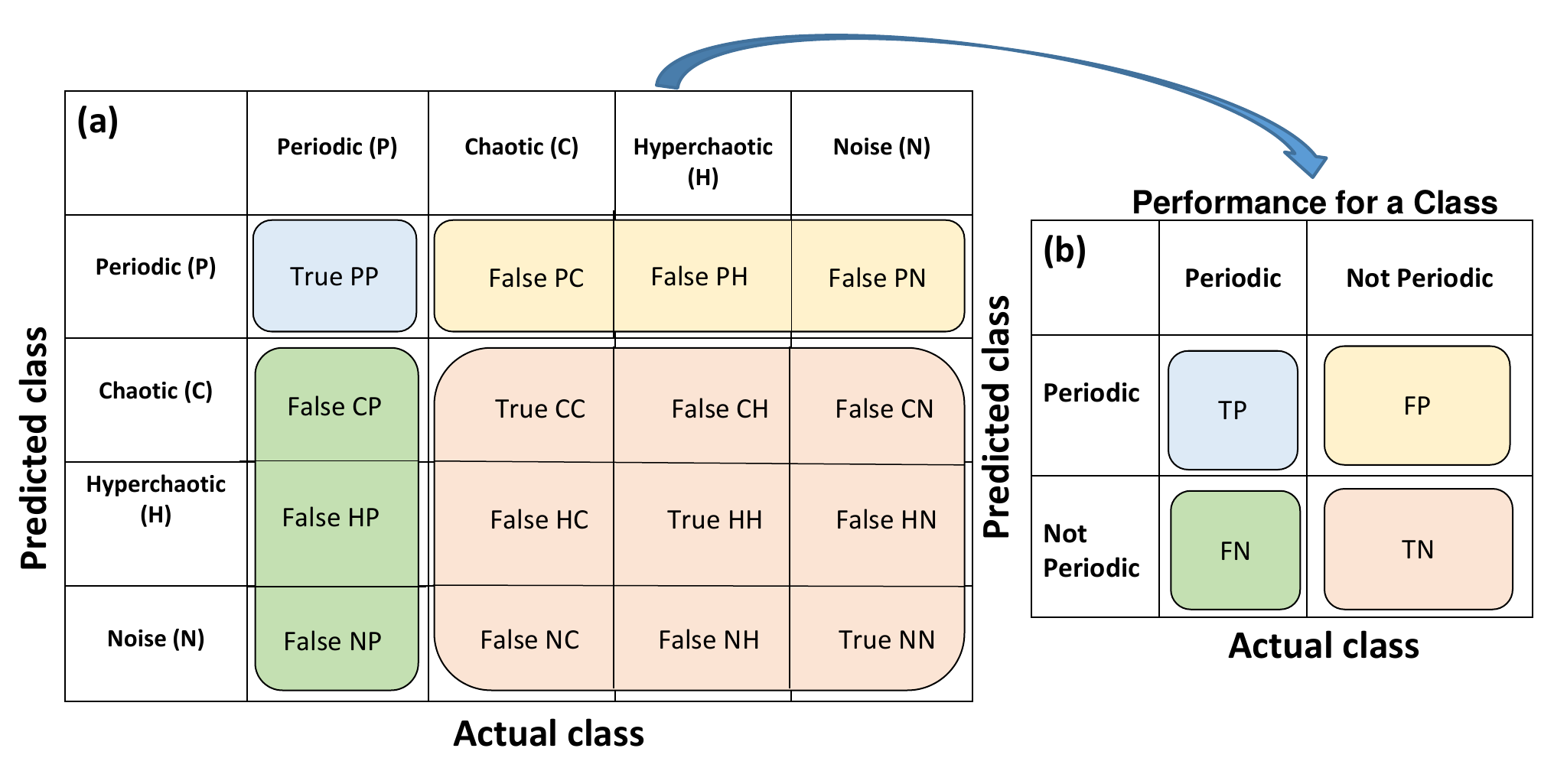}
\caption{\label{fig:CF} Confusion matrix for multi-class classification of time series data into Periodic, Chaotic, Hyperchaotic or Noise. To find the overall accuracy of the classifiers, we use the multi-class confusion matrix shown in the left panel (a), and to find performance for each class, we use the binary confusion matrix as shown in the right panel (b) for the periodic class for reference. }
\end{figure}
\begin{figure*}
  \centering
  \includegraphics[width=\textwidth]{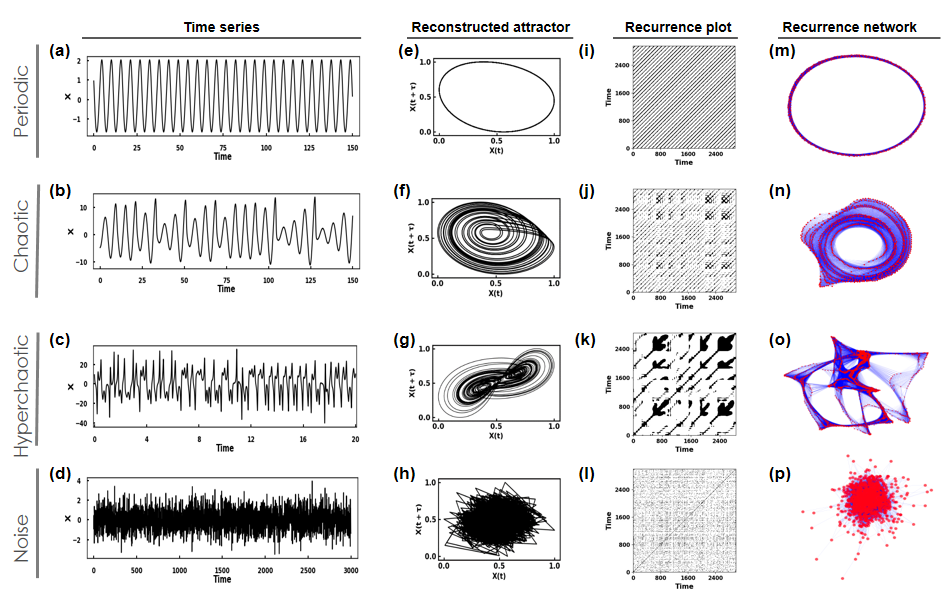}
  \caption{\label{TimeSeriesAndRP}Data sets generated for training and testing of the classification scheme.  The four types of nonlinear time series data (a)\textit{Periodic} time series obtained from R{\"o}ssler system with parameters $a=0.2$, $b=0.2$, $c=1$; (b) \textit{Chaotic} time series obtained from R{\"o}ssler system with parameters $a=0.2$, $b=0.2$, $c=7$; (c)\textit{Hyperchaotic} time series obtained from Chen system with parameters $a=35$, $b=4.9$, $c=25$, $d=5$, $e=35$ and $k=24$; and (d)White Noise obtained with Gaussian distribution with mean=0 and standard deviation=1. The re-constructed phase-space structure of the attractors (e-h), recurrence plots (i-l) and recurrence networks (m-p) corresponding to each time series} with red dots as nods and blue lines as edges. 
\end{figure*}
\subsection{\label{sec:level1}Machine learning approach to classification of time series}
 
In machine learning, the classification problem is related to identifying the category (class) of the new instance among the set of categories
(classes). When the categories contain more than two different classes,
 the problem is called a multi-class classification. 
The primary objective of multi-class classifiers is to assign an object to a specific class from a given set of classes based on the object's features and a training dataset.
In our case, the object of interest is a time series. We employ supervised machine learning methods to analyze datasets and build models capable of categorizing data into predefined and distinct classes. Specifically, we utilize three widely recognized machine learning algorithms\cite{MLAndTimeseries}: Logistic Regression (LR), Random Forest (RF) and Support Vector Machine (SVM). These algorithms play a vital role in supervised classification methods. In the following paragraphs, we provide brief descriptions of these algorithms, including the ranges of the input parameters. \\
\textbf{Logistic Regression (LR)}:
 In the one versus rest approach, each class in a dataset is treated as the positive class one at a time, while all other classes are grouped together as the negative class. This means that binary logistic regression models are built for each class individually.
When it comes to predicting on new data, the algorithm generates probability estimates for each class in the dataset. The class with the highest probability among these estimates is then chosen as the algorithm's prediction\cite{BinaryForMulti}. \\
\textbf{Random Forest (RF)}:
 It is a powerful tool in machine learning that combines many decision trees to make accurate predictions. The decision trees can be considered as a series of questions that lead to a conclusion. RF creates a "forest" of these trees \cite{RandomForest1}. 
Each tree operates on a random subset of data during training, contributing to the collective inference of the forest \cite{RandomForest1}. This ensemble nature enables RF to achieve superior prediction accuracy compared to individual trees, with errors diminishing and stabilizing as more trees are added. Hyperparameter tuning, facilitated by methods like 'GridSearchCV'\cite{GridSearch}, further optimizes RF models by adjusting parameters such as the number of estimators and maximum samples. Through meticulous hyperparameter tuning, RF maximizes its predictive capabilities, leveraging the collective knowledge of its decision tree ensemble.\\
\textbf{Support Vector Machine (SVM)}:
This is a widely used technique that performs intricate data transformation as determined by the chosen kernel function. These transformations are aligned to maximize separation boundaries between data points of different classes. We check for the regularization parameter $C$ values in the set $[0.1,1,10,100]$. In the case of utilizing the Radial Basis Function (RBF) kernel, an additional hyperparameter, $\gamma$ \cite{svm} must be specified before model training. The parameter $\gamma$  plays a crucial role in determining the curvature of the decision boundary. Commonly, $\gamma$ values like $[1, 0.1, 0.05, 0.01, 0.001]$ are considered, and the GridSearchCV is used to choose the best values for these parameters.\\ 
\subsection{\label{sec:level1}Performance analysis of the machine learning algorithms}

For estimating the performance of the LR, RF and SVM machine learning algorithms or classifiers, we apply \textit{cross validation} that involves dividing the dataset into $k$ equal sections, ensuring that the proportion of samples from each class is maintained within each fold. In our investigation, we adopt the widely used $10$-fold cross-validation scheme.\\
The performance\cite{PerClassAccuracy} of the classifiers are assessed in terms of sensitivity, specificity,  accuracy, precision, recall and F1-score to distinguish among all the four types of classes that may emerge in the time series.
 For the performance evaluation, in our study, first, we generate the confusion matrix to assess the model's performance collectively and comprehensively.
 In our study, first, we find a multi-class confusion matrix of $4 \times 4$ dimension to classify four patterns underlying the time series (FIG.\ref{fig:CF}(a)). Then, we extract a set of binary confusion matrices for the performance analysis for each class; for example, the performance of the periodic class is evaluated as shown in (FIG.\ref{fig:CF}(b)).
 The performance of binary confusion matrix corresponding to each class is determined in terms of accuracy, sensitivity, specificity, precision, and F1-score by counts of true positives (TP), false positives (FP), false negatives (FN), and true negatives (TN) \cite{AccuracyMetrics}. The mathematical formula to extract the performance measures for the classifiers for each class is as follows:\\
 
\noindent The accuracy score overall tells about the performance of the specific classifier, and it is defined as the number of instances for a particular class that are correctly predicted by the model and divided by the total number of instances of that class in the dataset.
\begin{equation}
\text{Accuracy} = \frac{\text{TP+TN}}{\text{TP+TN+FP+FN}}
\end{equation}
Sensitivity indicates out of all the actual positive cases, how many the model correctly identifies.  A high sensitivity value means the model is good at finding the most relevant positive cases in the data.
\begin{equation}
\text{Sensitivity} = \frac{\text{TP}}{\text{TP} + \text{FN}}
\end{equation}
Specificity estimates,  out of all the actual negative cases, the number that the model correctly identify as negative. Thus, it is the opposite of sensitivity (or recall), which focuses on detecting positive cases. High specificity means the model is good at correctly classifying true negatives, reducing the chances of mistakenly flagging something as a positive event when it is not.
\begin{equation}
\text{Specificity} = \frac{\text{TN}}{\text{TN} + \text{FP}}
\end{equation}
Precision gives the number of cases that are actually correct out of all the positive predictions made by the model. So a high precision value means that the model is good at making positive predictions that are highly likely to be true.
\begin{equation}
\text{Precision} = \frac{\text{TP}}{\text{TP} + \text{FP}}
\end{equation}
The F1 score is a way to find a good compromise between having a trustworthy friend (high precision) and a diligent detective (high recall). It considers both aspects of a model's performance.
 A high F1 score indicates the model balances between minimizing false positive errors (precision) and ensuring it doesn't miss actual positive cases (Sensitivity).
\begin{equation}
\text{F1 Score} = 2 \cdot \frac{\text{Precision} \cdot \text{Sensitivity}}{\text{Precision} + \text{Sensitivity}}
\end{equation}
 To measure the performance of the classifiers for multi-class classification, we find accuracy using a multi-class confusion matrix. Accuracy measures the performance of an algorithm irrespective of the class and it is defined as: 
\begin{equation}
\text{Accuracy} = \frac{\sum_{i=1}^{N} \text{TP}(C_i)}{\sum_{i=1}^{N} \sum_{j=1}^{N} C_{i,j}}
\end{equation}
where $N$ in the number of classes, TP$(C_i)$ is the number of correctly predicted instances for class $C_i$
and $C_{i,j}$ represents the element in the $i^{th}$ row and $j^{th}$ column of the multi-class confusion matrix.

\section{\label{sec:level2}Results}

\begin{figure*}
  \centering
  \includegraphics[width=\textwidth]{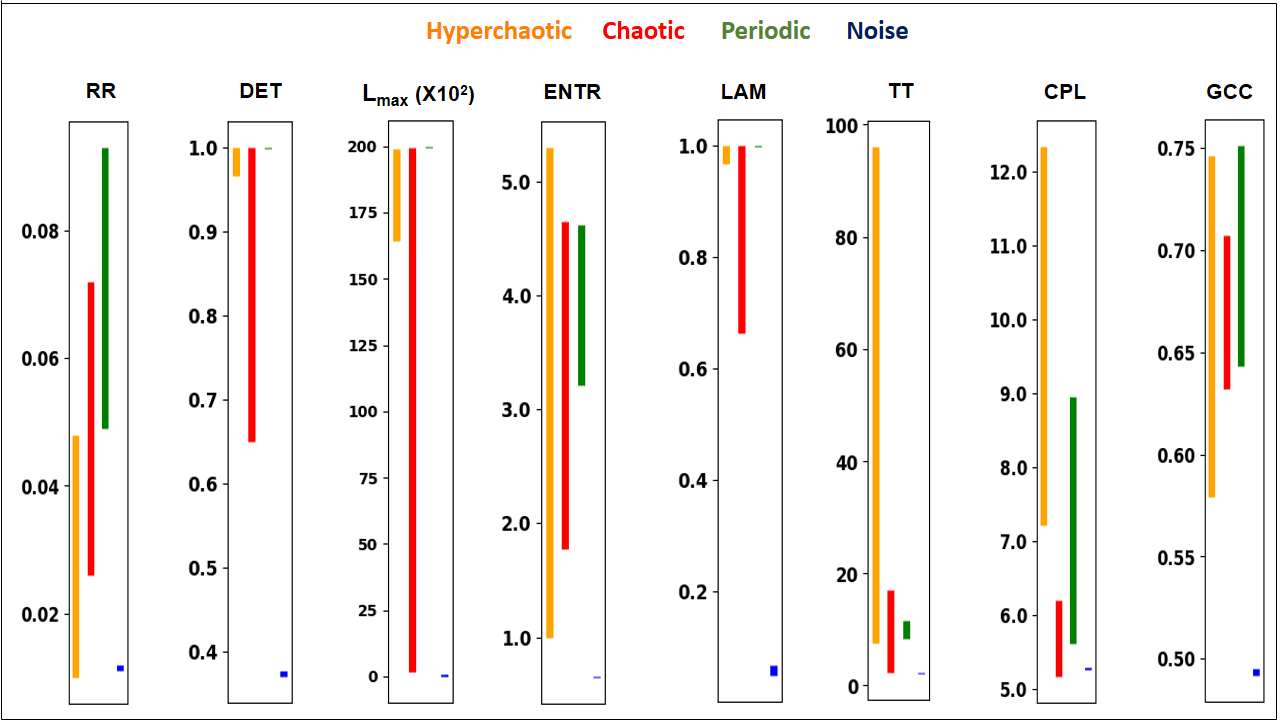}
  \caption{\label{Ranges of features} Ranges of the recurrence measures evaluated from the time series of dynamical states, periodic (green), chaotic (red), hyperchaotic (orange), and noisy data (blue) }
\end{figure*}

We begin by collecting time series data by simulating four standard dynamical models, namely, Lorenz, R{\"o}ssler, Duffing, and Chen systems (see table \ref{tab:table1_models}). We choose parameter values of these systems such that the time series shows three different dynamical states like periodic (FIG.\ref{TimeSeriesAndRP}a), chaotic (FIG.\ref{TimeSeriesAndRP}b), and hyperchaotic (FIG.\ref{TimeSeriesAndRP}c). In addition we consider white noise (FIG. \ref{TimeSeriesAndRP}d) which is generated using a Gaussian distribution with the mean zero. We reconstruct the phase space trajectories from each time series using appropriate time delay and embedding dimension. We then create its recurrence plot (FIG.\ref{TimeSeriesAndRP}i,j,k,l)) and recurrence network (FIG. \ref{TimeSeriesAndRP}m,n,o,p). Further, we extract various recurrence measures through recurrence quantification analysis of recurrence plot and extract average network properties from the recurrence network(FIG.\ref{Illustrative}e). We use these measures as input features for the machine learning algorithms (FIG.\ref{Illustrative}f).
 The results of our analysis for the recurrence measures, as depicted in (FIG.\ref{Ranges of features}) at recurrence threshold $\epsilon_2$, reveal interesting observations. When all the feature values are at the lower end of their respective ranges, it suggests that the time series may be noisy. Conversely, suppose the feature values are concentrated towards the upper end, especially for measures that quantify diagonal lines, such as DET, $L_{\text{max}}$, and LAM we can say that the time series is more likely to exhibit periodic behaviour.

\begin{figure*}
  \centering
  \includegraphics[width=\textwidth]{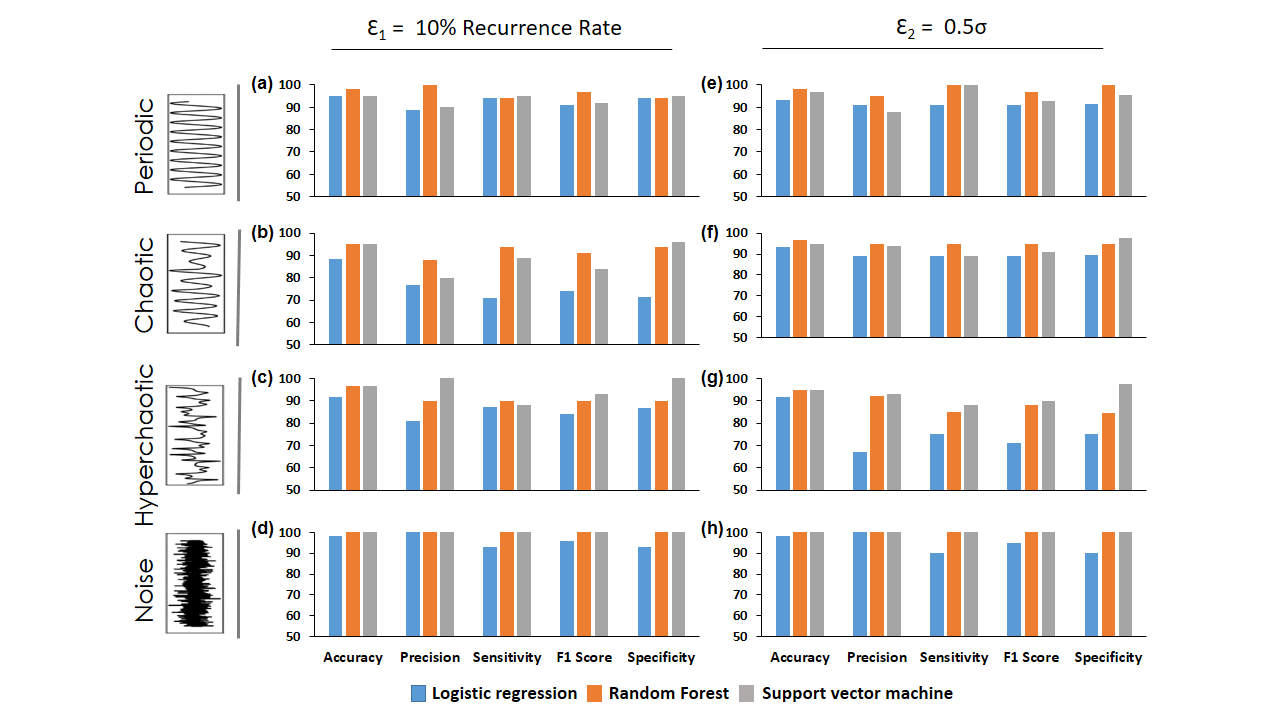}
  \caption{\label{performance_bar_graphs}Performance analysis of the three machine learning algorithms, namely Logistic Regression (blue), Random Forest (orange), and Support Vector Machine (grey). The left panel corresponds to  choice recurrence threshold $\epsilon_1$ and the right panel corresponds to $\epsilon_2$. The performance of each algorithm is evaluated in terms of accuracy, sensitivity, specificity, precision, and F1 score for all the classes of patterns. We can see the performance evolution of all machine learning algorithms for periodic patterns (a,e), chaotic patterns (b,f), hyperchaotic patterns (c,g), and for noise patterns (d,h).}
\end{figure*}
\begin{figure*}
  \centering
  \includegraphics[width=\textwidth]{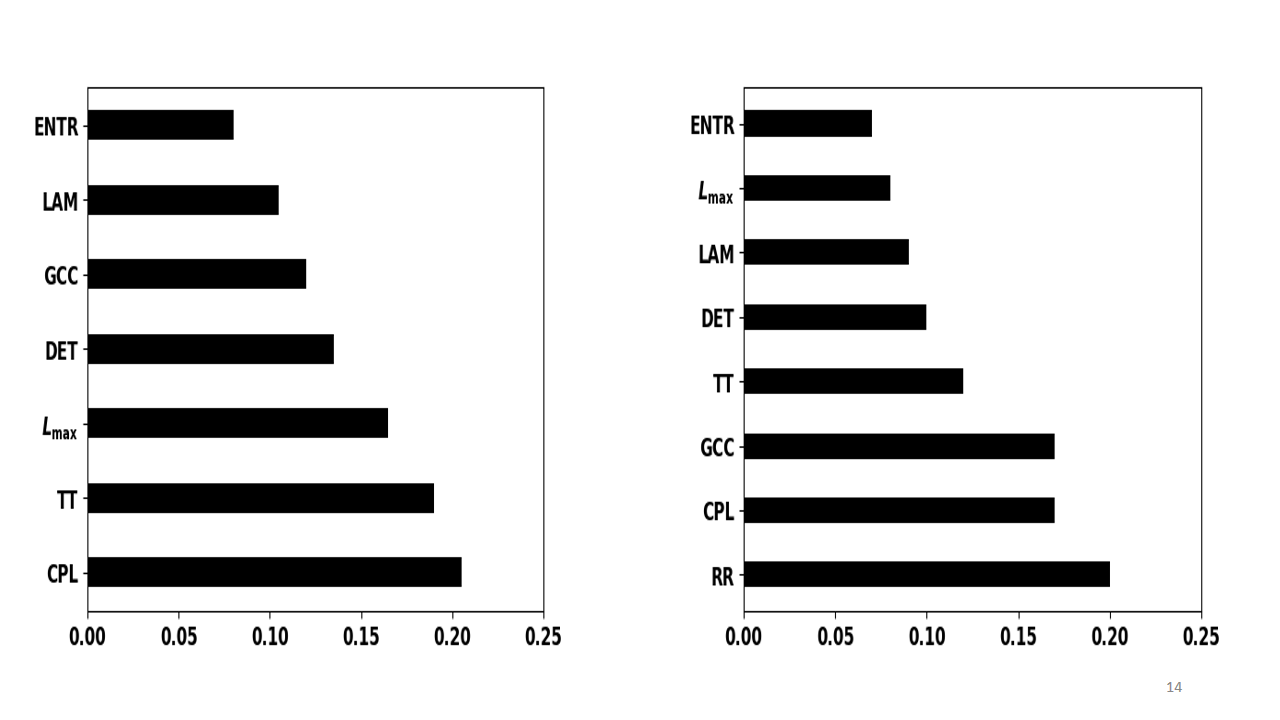}
  \caption{\label{feature_importance} Importance of input features- recurrence measures to detect the actual behaviour of nonlinear time series. The analysis uses the RF algorithm on the data sets generated at two recurrence thresholds $\epsilon_1$ (Left) 
 and $\epsilon_2$ (right). We observe that the features that measure the density of the closest recurrence points, like recurrence rate and average path length, are more important than the features that focus only on the measurement of the diagonal lines, like entropy and $L_{\text{max}}$.}
\end{figure*}
However, discerning chaotic and hyperchaotic patterns require exploring permutations of these features. This challenge is evident in FIG.\ref{Ranges of features}, indicating that it is hard to classify these patterns based solely on the feature ranges. We also observe similar results when recurrence threshold is $\epsilon_1$.
Therefore, we use machine learning algorithms to enhance the classification capabilities for dynamic behaviour. We apply the already mentioned machine-learning algorithms, each following different fundamental approaches, such as LR, RF and SVM.
 For the data sets, we consider seventy-five time series corresponding to each class, resulting in three hundred time series for all four classes. We take twenty-five time series from each of R{\"o}ssler, Lorenz, and Duffing oscillators for the periodic and chaotic class. For hyperchaotic class we take 25 time series from each of the Chen , 4D R{\"o}ssler and 4D Lorenz systems, resulting in a total of 75 time series. We divide the total time series into an $80:20$ ratio for training and testing purposes while ensuring that the number of time series corresponding to each class is equal, thus creating balanced data sets.
We check the performance of machine learning algorithms in terms of accuracy, sensitivity, precision, F1 score, and specificity for each class. We also find the accuracy using multi-class confusion matrix to check the performance of multi-class classifiers. 
We find that LR, RF and SVM algorithms give the accuracy  $86.67\%$, $95\%$, and $93.3\%$, respectively, irrespective of the recurrence thresholds $\epsilon_1$ and $\epsilon_2$ (see table \ref{tab:overall accuracy}).
 Therefore, we can conclude that  RF is a  better classifiers than LR and SVM for the multi-class classification problems related to classifying the states of time series among periodic, chaotic, hyperchaotic and white noise. 

 By analysing the performance of the machine learning algorithms for each class, our results reveal that the RF classifier achieved the highest accuracy for each class across both thresholds, outperforming the other two algorithms considered. The accuracy of both RF and SVM algorithms decreased as the complexity of the data increased, from periodic to chaotic and hyperchaotic classes. However, for the noisy class (Fig. \ref{performance_bar_graphs}d) – characterized by features concentrated within a lower concentrated range (Fig. \ref{Ranges of features}) – both RF and SVM achieved perfect accuracy of 100\%. In terms of overall performance balance between precision and sensitivity, as measured by F1 score, RF emerged as the better choice for periodic and chaotic classes (\ref{performance_bar_graphs}a,b). For the hyperchaotic class, however, SVM demonstrated slightly superior performance. Additionally, SVM's high specificity value for the hyperchaotic class reinforces its ability to accurately identify true negatives (data points that don't belong to the class) (\ref{performance_bar_graphs}c). \\
 
\noindent\textbf{\textit{Time required for the training data sets}}: We also examine the training time for all ML algorithms corresponding to both recurrence thresholds. According to the table \ref{tab:time}, we observe that RF takes the longer time than LR, and SVM requires the least amount of time to train the same data, regardless of the threshold value.\\

\noindent\textbf{\textit{Tolerance to noise in data}}:
To investigate how white noise contamination in the time series affects our analysis, we introduce varying proportions of white noise into the randomly chosen 30 chaotic time series. Our findings are presented in the table \ref{tab:table2_noiseAddition} for noise addition at four different levels: 5\% (Signal-to-Noise Ratio, SNR of $20$), $6\%$ (SNR of $16.67$), $7\%$ (SNR of $14.29$), and $8\%$ (SNR of 12.5)\cite{jacob2016characterization}. We use SVM and RF algorithms  since both of them show higher accuracy than LR for chaotic class.  We find that both SVM and RF algorithm  can tolerate noise up to $5\%$. We can say that up to $5\%$ noise addition, the original chaotic behaviour is predictable. However, accuracy drops signifcanly faster for SVM as compared to RF algorithm as we increase the percentage of white noise.\\

\noindent\textbf{\textit{Importance of features}}:
 The next step in our analysis is to determine each feature's relative importance to classify the patterns that emerge in nonlinear time series. Assessing the significance of the various features plays a pivotal role in ranking features, deciphering data, and gaining insights into the underlying phenomena in various practical scenarios. We arrive at the importance of features using the RF classifier. Figure \ref{feature_importance}(left) shows that average path length is the most influential feature, which contributes more than the other features in the multi-class classification at recurrence threshold $\epsilon_1$. However, from figure \ref{feature_importance}(right), we find that at $\epsilon_2$, the recurrence rate is a most influential parameter. Both these features are related to the density of the close points; for instance, the recurrence rate is the density of the recurrence points, and the average path length is the average of the shortest paths. Therefore, we can say that for the multi-class classification of patterns of time series, the features that measure the density of the closest recurrence points are the most important ones and the features that focus only on the measurement of the diagonal lines, like entropy and $L_{\text{max}}$, are comparatively less important.\\ 
\begin{table}
\caption{\label{tab:overall accuracy}Accuracy scores of the three ML algorithms calculated using multi-class confusion matrices for the time series data using two recurrence thresholds $\epsilon_1$ and $\epsilon_2$.}
\begin{ruledtabular}
\begin{tabular}{ccc}
Algorithms & $\epsilon_1$ & $ \epsilon_2$\\
\hline
LR&86.7&86.7\\
RF&95.0&95.0\\
SVM&93.3&93.3\\
\end{tabular}
\end{ruledtabular}
\end{table}
\begin{table}
\caption{\label{tab:time}Time(in seconds) taken to train the three ML algorithms using the time series data for two recurrence thresholds $\epsilon_1$ and $\epsilon_2$. }
\begin{ruledtabular}
\begin{tabular}{ccc}
Algorithms & $ \epsilon_1$ & $ \epsilon_2$\\
\hline
LR&0.052&0.054\\
RF&0.143&0.137\\
SVM&0.024&0.026\\
\end{tabular}
\end{ruledtabular}
\end{table}
\begin{table}
\caption{\label{tab:table2_noiseAddition}Accuracy scores of the ML algorithms computed for detecting the chaotic class within a set of chaotic time series, which are contaminated by white noise.}
\begin{ruledtabular}
\begin{tabular}{cccc}
Percentage of white noise&RF&SVM\\
\hline
5&98.3&98.3\\
6&76.7&43.3&\\
7&63.3&14.3\\
8&36.7&3.3\\
10&6.7&0.0\\
\end{tabular}
\end{ruledtabular}
\end{table}\\
\textbf{\textit{Prediction of the dynamical state from real data}}:
\begin{figure*}
  \centering
  \includegraphics[width=\textwidth]{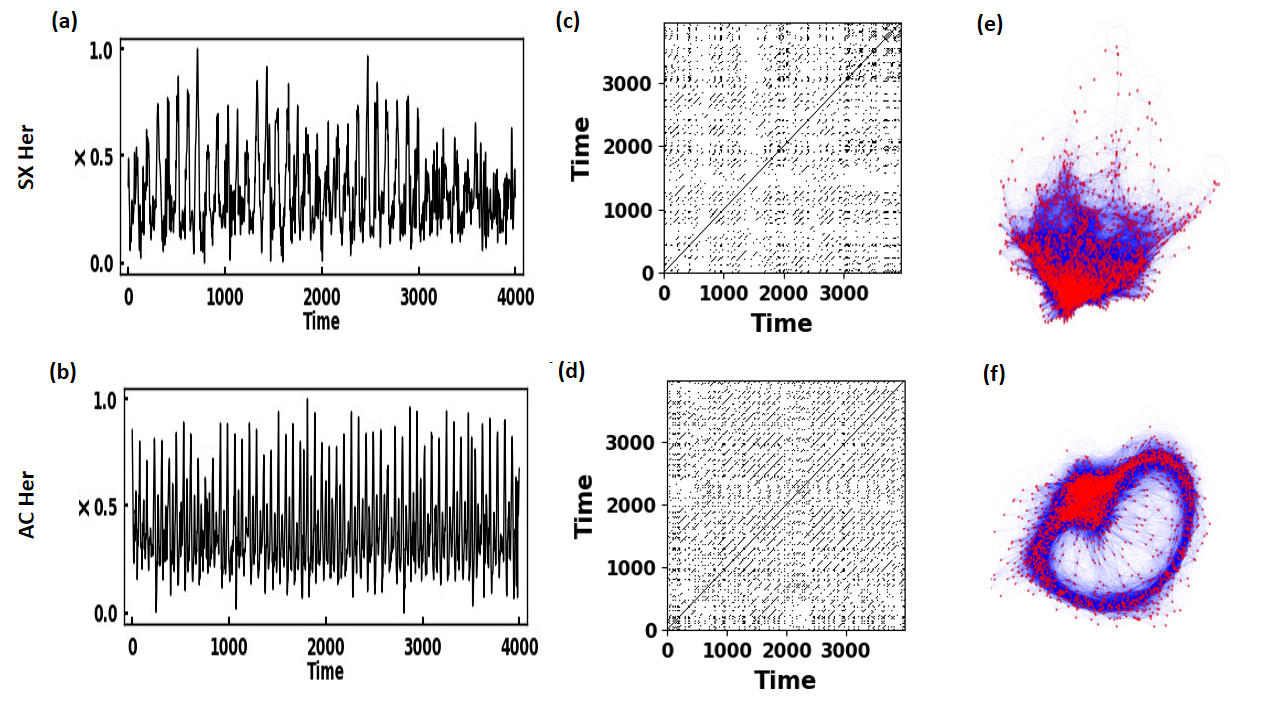}
  \caption{\label{real data} Normalized time series (a,b), recurrence plots(c,d) and recurrence networks (e,f) corresponding to the variable stars SX Her and AC Her }
\end{figure*}
As an illustration of our study to real world data sets, we apply the same multi-class classification procedure on time series data of two variable stars, SX Her (FIG \ref{real data}(a)) and AC Her (FIG \ref{real data}(b)). The data for these variable stars is taken from the AAVSO database (https://www.aavso.org/data-download). For preprocessing of the data, we initially compute the three-day averages and employ the cubic spline with a smoothing parameter to smooth and interpolate the data \cite{SXHer, ACHer}. We then embed the data in a 4-dimensional phase space and generate recurrence plots (FIG \ref{real data}(c,d)) and recurrence networks (FIG \ref{real data}(e,f)) for SX Her and AC Her. Since our analysis with synthetic data showed RF as the best classifier, we apply the same trained RF to the data from these stars.
The dynamics of these stars are reported as chaotic \cite{SXHer,ACHer} and we observe that both their data are predicted correctly as chaotic by the trained RF with recurrence threshold $\epsilon_1$.  However, for the recurrence threshold $\epsilon_2$, the time series corresponding to AC Her is predicted as chaotic, but SX Her is predicted as noise. 
When we extract the features corresponding to their time series, we observe that the values of all the features except global GCC are lying inside the range of features for the chaotic regimes when the recurrence threshold is set at $\epsilon_1$ (see table \ref{tab:table3}). Therefore, even though the GCC feature value is outside the chaotic range, the RF classifier still accurately predicts its class.
In this context, we make the following observations. Firstly, the GCC values for these stars are not significantly deviated from the range for chaotic data. Secondly, the GCC feature is less crucial for classification than determinism, $L_{\text{max}}$, trapping time, and average path length (see Fig. \ref{feature_importance}).
When the recurrence threshold is set at $\epsilon_2$, the values of all the features except GCC are within the range of chaotic, which is why the class of the AC HER star is predicted correctly. However, for the star SX Her, the values of the features ENTR, GCC are outside the range of the values of trained data sets, including the important feature RR. This may be the reason why its class is not predicted accurately.  \\

\begin{table}
\caption{\label{tab:table3}Feature values from data of variable stars  compared with the range from chaotic data with threshold as $\epsilon_1$}
\begin{ruledtabular}
\begin{tabular}{cccc}
Feature&Chaotic range&SX Her&AC Her\\
\hline
DET&(0.813 , 1.000)&0.961&0.968\\
$L_{\text{max}}$
&(175 , 19965)&2558&1761\\
ENTR&(1.668 , 4.845)&2.072&2.278\\
LAM&(0.921 , 1.000)&0.983&0.984\\
TT&(3.088 , 33.426)&5.989&5.688\\
L&(3.065 , 4.655)&3.133&3.287\\
GCC&(0.650 , 0.732)&0.619&0.648\\

\end{tabular}
\end{ruledtabular}
\end{table}
\begin{table}
\caption{\label{tab:table3}Feature values from data of variable stars  compared with the range from chaotic data with threshold as $\epsilon_2$}
\begin{ruledtabular}
\begin{tabular}{cccc}
Feature&Chaotic range&SX Her&AC Her\\
\hline
RR&(0.026 , 0.072)& 0.024&0.031\\
DET&(0.651 , 1.000)&0.814&0.916\\
$L_{\text{max}}$
&(153 , 19959)&375&334\\
ENTR&(1.774 , 4.646)&1.346&1.792\\
LAM&(0.663 , 1.000)&0.916&0.942\\
TT&(2.302 , 16.963)&3.588&3.838\\
L&(5.177 , 6.280)&5.682&5.497\\
GCC&(0.632 , 0.707)&0.547&0.575\\

\end{tabular}
\end{ruledtabular}
\end{table}

\noindent\textbf{\textit{Dynamical states from data of discrete systems}}:
\begin{table*}
\caption{\label{tab:table1_models_discrete} Details of discrete dynamical systems and their parameter values used to generate synthetic data with dynamical states of periodic and chaotic behaviour}
\begin{ruledtabular}
\begin{tabular}{p{1.5cm}p{3.5cm}p{3.5cm}p{2.5cm}}
 Time Series Generator & Dynamical Equations & Parameter Values & Dynamical Behavior\\\hline
Logistic map \cite{LogisticMap} & 
    $\begin{aligned}
       x_{n+1} = rx_n(1 - x_n)\\
    \end{aligned}$ & $ 
    \begin{tabular}[t]{@{}l@{}}
    $r \in (3.57, 4)$ \\
    $r \in (3, 3.5699)$ 
    \end{tabular}  $ & 
    $\begin{aligned} 
        \text{Chaotic} \\
        \text{Periodic}
    \end{aligned}$ \\\\ \hline 
H\'{e}non map \cite{Henonl} & 
    $\begin{aligned}
       x_{n+1} &= y_n + 1 - ax_n^2 \\
       y_{n+1} &= bx_n\\
    \end{aligned}$ & 
    $\begin{aligned} 
        b &= 0.4, \quad
        a \in (1.4272,1.5) \\
        b &= 0.4, \quad
        a \in (0.2, 0.9) 
    \end{aligned}$ &
    $\begin{aligned} 
        \text{Chaotic} \\
        \text{Periodic}
    \end{aligned}$ \\\\ \hline
H\'{e}non-logistic map(2D-HLM)\cite{2DHLM} & 
    $\begin{aligned}
       x_{n+1} &= a+by_n(1 - cy_n)-ax_n^2 \\
       y_{n+1} &= c^2y_n(1-by_n)\\
    \end{aligned}$ & 
    $\begin{aligned} 
        a &= 0.895, \quad
        c &= 2, \quad
        b &\in (1 , 4) \\
        a &= 0.895, \quad
        b &= 8, \quad
        c &\in (1.74,1.88) \\ 
    \end{aligned}$ &
    $\begin{aligned} 
        \text{Chaotic} \\
        \text{Periodic}
    \end{aligned}$ \\\\ \hline 
\end{tabular} 
\end{ruledtabular}
\end{table*}
We now extend our analysis to classify the dynamical states from time series data of discrete systems.  We generate periodic and chaotic time series using the Logistic Map\cite{LogisticMap}, Hénon Map \cite{Henonl} and the 2D Hénon-Logistic Map (2D-HLM) \cite{2DHLM} (see Table \ref{tab:table1_models_discrete}). Our dataset comprises 20 time series from each system for both chaotic and periodic classes, resulting in a total of 60 chaotic and 60 periodic time series. As reported in literature \cite{LogisticMapRP, HenonReconstruction} we get the recurrence plots and recurrence networks from the reconstructed attractors, with a recurrence threshold of 0.05 in standard deviation units \cite{DisThresh}. The ranges of features in this case differ from those of continuous systems and hence the algorithms are trained  using the new set of  features to classify the dynamical states from data of  discrete systems. The accuracy achieved by the three machine learning algorithms, LR, RF, and SVM, for the discrete dataset is presented in the Table  \ref{tab:discrete accuracy}.
\begin{table}.
\caption{\label{tab:discrete accuracy}Accuracy scores of the three ML algorithms calculated using binary class confusion matrices for the time series data of discrete systems }
\begin{ruledtabular}
\begin{tabular}{cc}
Algorithms & Accuracy \\
\hline
LR&95.8\\
RF&95.8\\
SVM&95.8\\
\end{tabular}
\end{ruledtabular}
\end{table}

\section{Conclusion}
We report a detailed and comprehensive study on the application of three supervised machine learning algorithms, LR, RF, and SVM, to classify the multi-classes of nonlinear time series, such as periodic, chaotic, hyperchaotic, and noise. The required input features for the machine learning algorithms are extracted using recurrence quantification analysis of recurrence plots and using average network properties of recurrence networks. We collect synthetic time series data by simulating four standard dynamical models of R{\"o}ssler, Lorenz, Chen, and Duffing oscillator systems and time series generated from the Gaussian distribution. We analyse the performance of the three classifiers on the data sets generated at two recurrence thresholds. We observe that irrespective of the recurrence threshold, all the classifiers effectively predict the nonlinear time series' dynamical behaviour and noisy patterns. Among the three classifiers, the RF algorithm exhibits the highest accuracy in identifying dynamical states and noise behaviour within nonlinear time series. SVM follows RF as the second-best classifier in this regard.

We extract feature importance using the RF method for the data sets obtained using two different recurrence thresholds $\epsilon_1$ and $\epsilon_2$. Our study indicates that among all the features we used, the features related to the density of the recurrence points or the closeness of points, like recurrence rate, average path length and trapping time, are the most effective in predicting the class of a given time series correctly.
 We also observe that the classifiers can tolerate up to $5\%$ noise in the chaotic data set. This means the chaotic state of the time series, which includes noise up to $5\%$, can be correctly identified using machine learning algorithms.
 
 Further, we apply the same procedure of the multi-class classification to predict the dynamical state of the real data of the time series from two variable stars, SX Her and AC Her, and successfully predict the actual dynamical state of both stars at the recurrence threshold value $\epsilon_1$. Furthermore, we could train the algorithms and successfully detect periodic and chaotic dynamical states emerging in the discrete systems. Our analysis reveals that all three machine learning algorithms predict these dynamical states with an accuracy of $95.8\%$. This demonstrates the effectiveness of our approach in accurately classifying the dynamical states from the data obtained from both continuous and discrete dynamical systems. 
 In future, our aim is to enhance the integration of algorithms designed for both continuous and discrete systems. This involves expanding the training dataset to incorporate additional systems with appropriate recurrence thresholds, thereby improving the generalization and robustness of the algorithms. We hypothesize that these refined models will not only classify systems as continuous or discrete but will also provide insights into their specific dynamics, such as periodic, chaotic, hyperchaotic, and noisy behavior.Then the trained algorithm can be used to test time series data derived from realistic systems, including domains such as astronomy, oceanography, meteorology, and satellite operations. This will contribute to assessing the applicability of our models in real-world scenarios where traditional models may fall short.
Additionally, we can make a GUI using our algorithm so that researchers and practitioners can seek quick and accurate insights into the dynamics of their time series data.
Finally, our ongoing efforts involve exploring the feasibility of classifying dynamical states based solely on visual representations, such as images derived from recurrence plots. This avenue of research has the potential to simplify the process and broaden the applicability of our methodology.

  \begin{acknowledgements}
 D.T. and C.M. would like to acknowledge the INSPIRE-Faculty grant (code IFA19-PH248) of the Department of Science and Technology, India, for financial support.
\end{acknowledgements}
\section*{Data Availability Statement}
Data and the code can be made available upon request to the authors.

\bibliographystyle{unsrt}
\bibliography{Timeseries}

\end{document}